\pdfoutput=1

\documentclass[a4paper,fleqn, 11pt]{cas-sc}

\usepackage[numbers]{natbib}
\usepackage[utf8]{inputenc}
\usepackage{booktabs}
\usepackage{float}
\floatstyle{plaintop}
\restylefloat{table}

\usepackage[center]{caption}

\usepackage{array}
\newcolumntype{P}[1]{>{\centering\arraybackslash}p{#1}}

\usepackage{lipsum}
\usepackage{setspace}

\usepackage{adjustbox}

\usepackage{lineno}

\usepackage{booktabs}
\usepackage{url}
\usepackage{tabularx}
\usepackage{multirow}
\usepackage{amsmath,amssymb,amsfonts}
\usepackage{algorithmic}
\usepackage{graphicx}
\usepackage{textcomp}
\usepackage{xcolor}

\DeclareUnicodeCharacter{2212}{-}
\begin{document}

\let\WriteBookmarks\relax
\def\floatpagepagefraction{1}
\def\textpagefraction{.001}
\shorttitle{}
\shortauthors{Ekiz et~al.}

\title [mode = title]{ Long Short-Term Network Based Unobtrusive Perceived Workload Monitoring with Consumer Grade Smartwatches in the Wild}                      

\tnotetext[1]{  This work is supported by the Turkish Directorate of Strategy and Budget under the TAM Project number DPT2007K120610.}


\author[1]{Deniz Ekiz}[type=editor,
                        orcid=0000-0002-8130-3841
                        ]
\cormark[1]
\ead{deniz.ekiz@boun.edu.tr}


\address[1]{Department of Computer Engineering, Boğaziçi University, Istanbul, Turkey}

\author[1]{Yekta Said Can}[%
orcid=0000-0002-6614-0183
   ]


\author[1]{Cem Ersoy}[%
orcid=0000-0001-7632-7067
]

\cortext[cor1]{Corresponding author}


\begin{abstract}
Continuous high perceived workload has a negative impact on the individual's well-being. Prior works focused on detecting the workload with medical-grade wearable systems in the restricted settings, and the effect of applying deep learning techniques for perceived workload detection in the wild settings is not investigated. We present an unobtrusive, comfortable, pervasive and affordable Long Short-Term Memory Network based continuous workload monitoring system based on a smartwatch application that monitors the perceived workload of individuals in the wild. We make use of modern consumer-grade smartwatches. We have recorded physiological data from daily life with perceived workload questionnaires from subjects in their real-life environments over a month. The model was trained and evaluated with the daily-life physiological data coming from different days which makes it robust to daily changes in the heart rate variability, that we use with accelerometer features to asses low and high workload.  Our system has the capability of removing motion-related artifacts and detecting perceived workload by using traditional and deep classifiers. We discussed the problems related to in the wild applications with the consumer-grade smartwatches. We showed that Long Short-Term Memory Network outperforms traditional classifiers on discrimination of low and high workload with smartwatches in the wild.
\end{abstract}




\maketitle

\section{Introduction}
 \begin{figure} 
\centering
\includegraphics[width= 0.54\columnwidth]{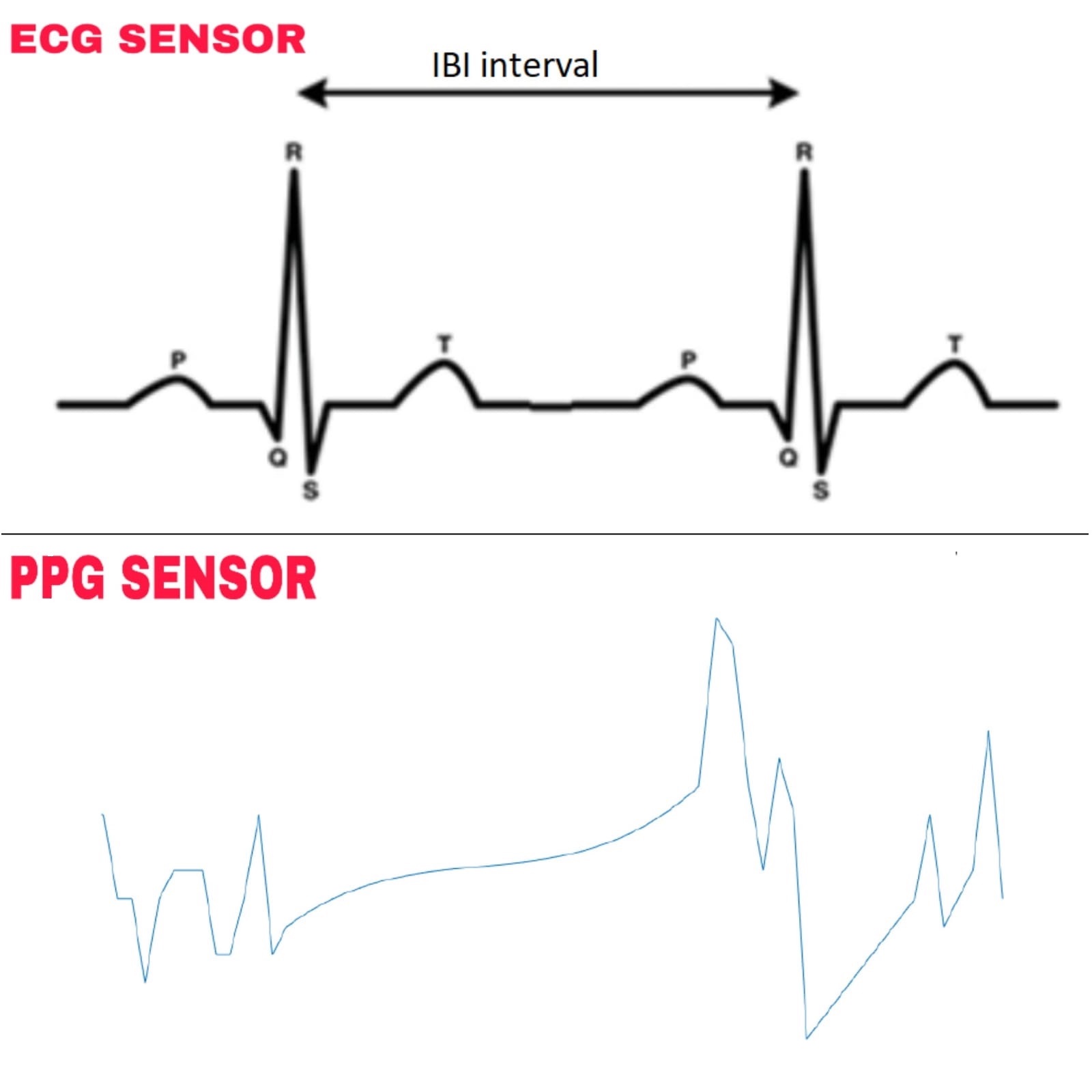}
\caption{The difference between the IBI data obtained from an ECG sensor and a PPG sensor. ECG data has a structure which composes of P, Q, R, S, T data points \cite{Poungponsri2013AnAF} whereas we can only obtain peaks from PPG data for capturing heartbeats. }
\label{diff_IBI_PPG}
\end{figure}
Ongoing high levels of perceived workload can have a negative impact on the well-being and health \cite{stressworkload}. Recent research has revealed concerning rates of anxiety and depression among university students \cite{Bayram2008}. Every individual can perceive a task with different levels of workload. For example, for a Computer Science student, writing a program to compute Fibonacci numbers may be easy, but it would be hard for a student from the philosophy department.  Traditionally, perceived workload can be assessed with self-reports. However, collecting these reports can be a challenging task.  In recent years, researchers experimented with wearable sensor technologies for identifying high workload. However, many of the investigated technologies were high-end sensor systems like full-lead  electrocardiography (ECG) \cite{RAHHAL2016340}  and  electroencephalography  (EEG) \cite{Fan2018}, which do not provide high-comfort.  Consumer-grade out-of-the-box smartwatches and smartbands can be used to monitor the perceived workload in daily life because they are unobtrusive, comfortable to wear and cost-efficient. Modern smartwatches are promising candidates for daily life perceived workload detection since they are equipped with built-in sensors including heart rate monitor, accelerometer and gyroscope. The photoplethysmography (PPG) based heart rate monitoring (HRM) units  of popular smartwatches such as Samsung Gear Series and Apple Watch provide less data than high-end sensor systems (sampling frequency 1000+ Hz vs less than 100 Hz) and more vulnerable to the artifacts which make the decision process of the proposed system harder (see Figure \ref{diff_IBI_PPG} for IBI data comparison of ECG and PPG sensors).  Generally, the number of valid RR intervals from full-lead ECG signals is about 99\%, whereas the same value for the smartwatches  can drop to 50\% \cite{Can2019} which makes the system developed in the laboratory to fail in the wild settings.

Perceived workload detection in the wild settings is harder than laboratory settings, due to the dependence on the self-reports, selection of unobtrusive devices and less information about the task that the user is engaging. More sophisticated approaches for daily life settings are required.

\indent Recently, more and more engineering problems have been solved with Recurrent Neural Networks (RNN)  since they are powerful tools with high performance. For automatic workload detection systems using physiological data, researchers needed a neural network that could take advantage of the sequential structure of the data. Long Short-Term Memory Network (LSTM) is a particular type of RNN. It was defined by
Hochreiter and Schmidhuber in 1997 \cite{hochreiter1997long} to alleviate the long-term dependency problem of RNN \cite{Alhagry2017}. Because of the vanishing/exploding gradient problem as a result of backpropagation through time in a standard RNN, a long sequential data could be hard to learn \cite{Alhagry2017}. In LSTM, instead of an RNN cell, a gated LSTM cell is used to cope with this problem. This makes LSTM suitable for sequential data classification. Chauchan et al. \cite{Chauhan2018} showed that LSTM can be implemented on a smartwatch and a smartphone, and works faster than the shallow SVM. 

\indent In this study, we improved the state of the art by combining two important features as shown in Table \ref{table:relatedwork}. First, our system tested and trained in the wild with self-reported questionnaires without any restriction of a certain task, and with the consumer-grade smartwatches which makes it scalable. Most of the previous works are tested on a certain type of task like the N-Back task less than a week. Detecting workload in the wild settings without restricting the task type is more challenging. Ground truth collection is solely based on questionnaires in the wild whereas the workload level is known at any moment in the restricted environments. When the unrestricted movements and resulting artifacts, low data quality of unobtrusive devices, subjectivity of the perceived workload obtained from questionnaires are taken into account, the performance of perceived workload detection systems in the wild is lower than the systems tested in the laboratory environments. 

Second,  to the best of our knowledge, our work is the first attempt that applies LSTM for daily life perceived workload detection with consumer-grade smartwatches. We compared the prediction performance of LSTM with a shallow network and the traditional machine learning algorithms on the perceived workload data coming from daily life. 

The rest of the paper is organized as follows: In Section \ref{chaptertwo}, we present the related work for automatic workload detection systems in daily life. In Section \ref{methodology}, we present the proposed smartwatch based workload detection platform. In Section 5, we explained the data collection procedure. In Section \ref{sec:results}, we provide the results of the proposed system. In Section \ref{discussion}, we evaluate the findings of the study and future work of the current research.

\section{Related Works}
\label{chaptertwo}

In the early workload studies with wearables, data from the different physiological modalities were collected in laboratory environments and over 90\% accuracy was achieved for detecting two levels of workload \cite{Fan2018}. However, researchers noticed that affect levels experienced in the laboratory environments are different when compared to the real-life situation affect levels \cite{PicardLab2016}. 
Cinaz et al. \cite{Cinaz:2013:MMW:2434601.2434680} developed an ECG chest belt based perceived workload detection system, they trained their model in the laboratory and applied in an office environment. They used Linear Discriminant Analysis (LDA), K-Nearest Neighbour (kNN) and SVM as the classifier. Their model recognized successfully five out of seven participant's perceived workload during one day. Fan et al. \cite{Fan2018} developed an EEG based perceived workload detection system and evaluated it in a Virtual Driving environment. They used kNN for the machine learning unit. Schaule et al. \cite{Schaule2018} developed a perceived workload detection scheme using a smartwatch (Microsoft Band 2) which is equipped with Galvanic Skin Response (GSR), HRV and Skin Temperature (ST) sensors with Naive Bayes, kNN and Random Forest classifiers. They evaluated their approach on the N-Back task with 10 participants for one day.  N-Back task is a protocol used in cognitive science studies to measure the working memory \cite{nback}.  They plan to add an "in-the-wild" application as future work. Munoz et al. \cite{Munoz2016} conducted an SDNN based (which is a feature of the HRV shown in the Table \ref{table:hrv_fatures}) statistical analysis by using LG Watch R, on 9 participants during three consecutive days. They applied the N-Back Task to alleviate the perceived workload.

These studies are conducted in semi-restricted environments  \cite{Munoz2016}, \cite{Cinaz:2013:MMW:2434601.2434680} and \cite{Schaule2018}. Although participants are in ambulatory settings and their movements are not restricted, they follow a pre-arranged schedule or program which makes labeling and assigning the ground truth easier. Also, they collected objective ground-truths along with the questionnaire, which makes their system more accurate but hard for further data collection due to the requirement of a dedicated laboratory settings. Workload detection in unrestricted real-life (daily life) is the most challenging problem since we rely only on personal self-reports as the ground truth. The classification accuracies are lower than those in the laboratory environments and there is a room for improvement in the performance of the daily life perceived workload detection research (see Table \ref{table:relatedwork}) \cite{Schaule2018}. Researchers strive hard to come up with ways to improve the performance of daily life workload detection studies.  As seen from the literature, a system that is designed for daily life settings is required. For the sake of the scalability, the only option to train this system would be based on self-reported questionnaires.

\begin{table}[]
  \caption{Comparison of our work with the previous works on workload detection.}
  \centering
  \label{table:relatedwork}
\resizebox{\columnwidth}{!}{
\begin{tabular}{p{3.5cm}P{1.6cm}P{2cm}P{1.5cm}P{1.7cm}P{1cm}P{1.3cm}P{1.6cm}P{1.3cm}}
\hline
Article         & Device           & Features                                                                           & Method                                                               & Environment     & Classes & Accuracy            &  Participants & Duration \\ \hline
Munoz et al. (2016) \cite{Munoz2016}     & LG Watch  R      & SDNN                                                                               & ANOVA                                                                & N-Back Task     & N/A               & N/A                 & 9           & 3 days   \\  \\ \hline
Fan et al. (2018) \cite{Fan2018}      & EEG              & Statistical, FD,  HOC & kNN                                                                 & Virtual Driving & 2                 & 82 \% recall         & 20          & 6 days   \\ \\ \hline
Cinaz et al. (2013) \cite{Cinaz:2013:MMW:2434601.2434680}    & ECG Chest Belt   & HRV                                                                                & LDA, kNN, SVM            & Office-Work     & 3                 & 71\%                & 7           & 1 day    \\\\ \hline
Schaule et al.  (2018) \cite{Schaule2018}  & Microsoft Band 2 & GSR, HRV, ST                                    & Random Forest, Naive Bayes SVM                & N-Back Task         & 2                 & 66\%*  & 10          & 1 day    \\\\ \hline
Our Work (2019) & Samsung Gear S2  & ACC, HRV                                                                            & LSTM, MLP, SVM Naive Bayes, kNN, Random Forest & In the wild     & 2                 & 70\%                & 17          & 1 month   \\  
\hline
\multicolumn{4}{l}{\textsuperscript{*}\footnotesize{Average accuracy is not reported.}}

\end{tabular}

}
\end{table}

 \begin{figure} 
\centering
\includegraphics[width= \columnwidth]{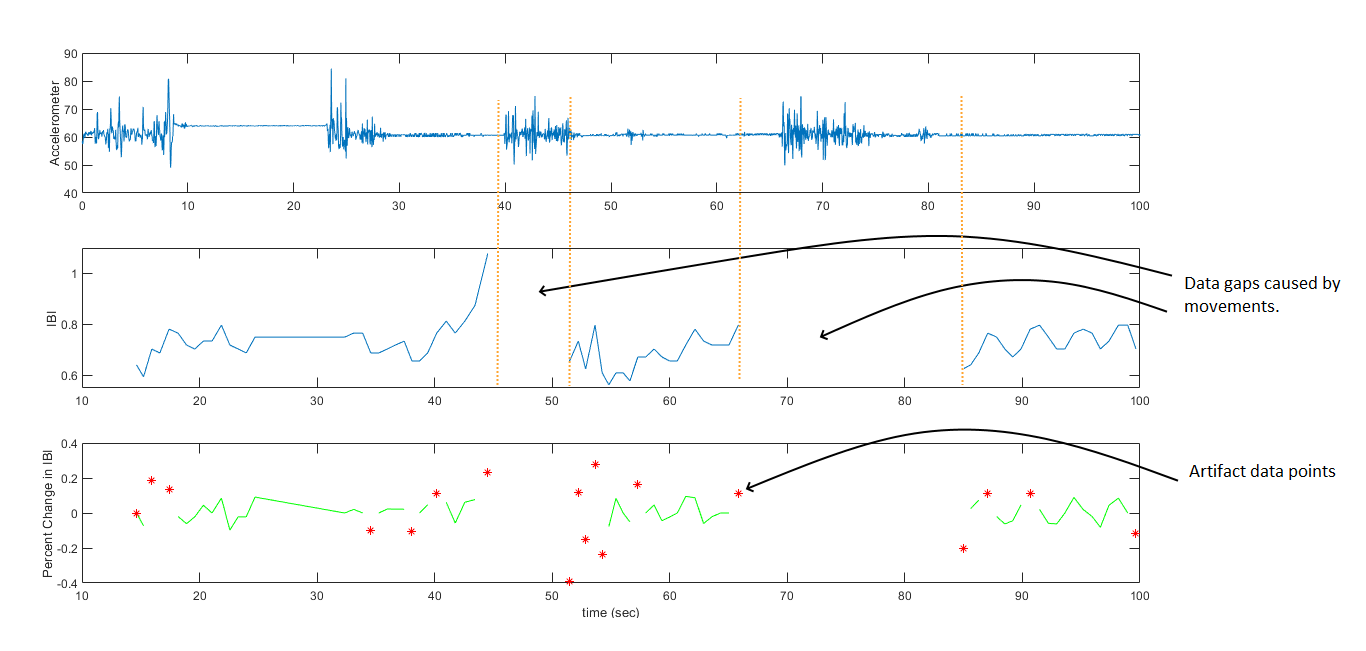}
\caption{Data gap and artifact problems caused by motions. When the physical activity (see Accelerometer data in the first subfigure) is increased above some threshold, data gaps have emerged. Furthermore, minor hand movements cause artifacts in data. In the last subfigure, we extracted percentage changes in the IBI data. We marked the artifacts which are more / less from the local average above 20 percent. }
\label{datagap}
\end{figure}

\section{Proposed System Description}
\label{methodology}
\begin{figure*}
\centering
\includegraphics[width=\textwidth]{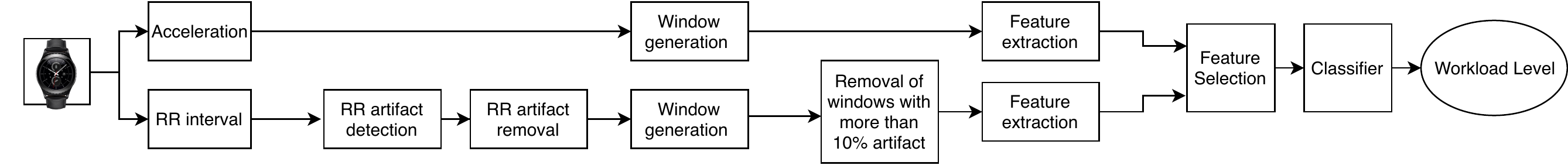}
\caption{The system diagram of the perceived workload monitoring system.}
\label{figurefors2}
\end{figure*}

In this study, we propose a smartwatch based perceived workload level detection system that works completely in the wild. This system monitors the user's self-reported perceived workload levels in their daily activities without creating any interruption or restriction. The only requirement to use this system is wearing a smartwatch. Daily-life environments create more challenges than controlled laboratory setups \cite{survey}. Artifacts caused by unrestricted movements and loosely worn smartwatches are among these problems. Extremely loose settings lead to artifacts due to lack of skin contact and the PPG sensor is highly affected by the motion artifacts (see Figure \ref{datagap}) \cite{motionppg}. Therefore, we applied artifact detection and removal techniques. After the artifacts are removed, we interpolated the missing data points. Next, the features in Table II are extracted from the sensory signals and fed to the machine learning algorithm for prediction. In order to use this system, pre-trained machine learning models are required.  LSTM classifiers were trained on the feature vectors with generated class labels. We further compared LSTM with a shallow neural network and traditional machine learning algorithms.  The overall diagram of the proposed system is presented in Figure \ref{figurefors2}.

\subsection{Smartwatch Framework}

\begin{table}[]
\caption{Heart rate variability and Acceleration features and their definitions.}
\begin{tabular} {@{}p{0.3\textwidth}p{0.62\textwidth}@{}}
\hline \textbf{Feature} & \textbf{Description} \\ \hline 
 \multicolumn{2}{c}{\textbf{Heart Rate Variability Features}}\\ \hline
Mean RR & Mean value of the RR intervals \\
SDNN & standard deviation of the inter-beat interval \\
RMSSD & Root mean square of the successive difference of the RR intervals \\
pNN50 & Percentage of the number of successive RR intervals varying more than 50ms from the previous interval \\
HRV triangular index & Total number of RR intervals divided by the height of the histogram of all RR intervals measured on a scale with bins of 1/128 s \\
TINN & Triangular interpolation of RR interval histogram \\
LF & Power in low-frequency band (0.04-0.15 Hz) \\
HF & Power in high-frequency band (0.15-0.4 Hz) \\
LF/HF & Ratio of LF-to-HF \\
VLF & Power in very low-frequency band (0.00-0.04 Hz) \\
SDSD  & Related standard deviation of successive RR interval differences 
\\
\hline
 \multicolumn{2}{c}{\textbf{Acceleration Features}}\\ \hline
Mean X & Mean acceleration over x axis \\
Mean Y & Mean acceleration over y axis \\
Mean Z  & Mean acceleration over z axis\\
Mean ACC MAG &  Mean acceleration over acceleration magnitude axis \\
STD X  & Standard Deviation of acceleration over x axis\\
STD Y  & Standard Deviation of acceleration over y axis\\
STD Z  & Standard Deviation of  acceleration over z axis\\
Energy & FFT energy over mean acceleration magnitude \\ 

\hline 

\\ \hline 
\end{tabular}%
\label{table:hrv_fatures}
\end{table}

\begin{figure}
    \centering
    \includegraphics[width=0.5\columnwidth]{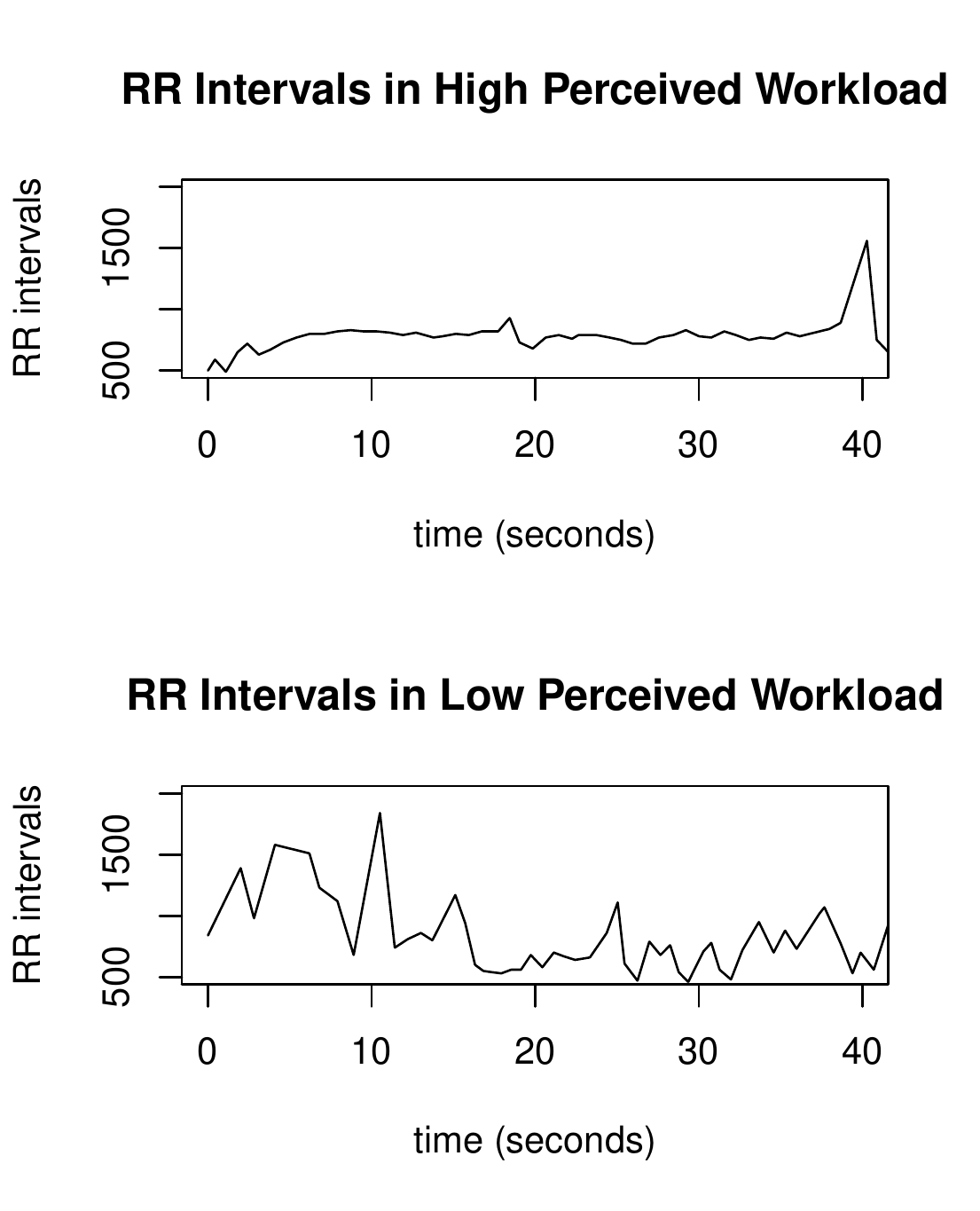}
    \caption{RR intervals in high and low perceived workload. The variability decreases in high perceived workload.}
    \label{fig:highwslow}
\end{figure}

 Samsung Gear S2 is equipped with the ambient light sensor, PPG sensor for heart rate monitoring, 3D inertial measurement unit with 3D accelerometer and gyroscope and pedometer. Samsung Gear Series run on the Tizen platform. We developed a data collection application for the Tizen Platform Wearable 2.3.2. The acceleration data collection application was developed in our previous works \cite{Ekiz2017}, \cite{hibit} and \cite{Can2019}. The sampling rate of the 3D accelerometer is 20 Hz. In order to compute RR intervals, the smartwatch samples PPG with 100 Hz and calculates each successive beat.   The retrieved RR intervals from the smartwatch application during low and high perceived workload can be seen in Figure \ref{fig:highwslow}. The decrease in HRV can be seen in the case of high workload, from the data gathered by the smartwatches.  We used Samsung Gear S2's haptic feedback functionality to inform participants that the session is finished and they should fill the questionnaire using their smartphones.
 Body and head movements can be used to detect the emotions and arousal level \cite{Ekman1965}. Montepare et al. \cite{Montepare1999} demonstrated video recordings of actors expressing emotions with body movements to the 82 younger and older adults. The face of actors are blurred, thus understanding the facial expression from the video recording is not possible. All videos were silent. Participants correctly identified the emotions expressed by actors in video recordings. From the accelerometer, body movements can be automatically recognised. HRV can be affected by movements also, therefore usage of acceleration along with HRV is required.

\subsection{Preprocessing of Physiological Signals}
\label{preprocessing}

All of the preprocessing codes are implemented in MATLAB \cite{matlab}. We developed the preprocessing and feature extraction modules in our previous work \cite{Can2019}. RR intervals are filtered for artifacts by calculating the difference of successive RR intervals. Acceleration and RR signals are divided into 2 minutes long sliding windows with 50\% overlap \cite{Cinaz:2013:MMW:2434601.2434680} , \cite{gjoreski2017monitoring}. We calculated local averages that took 10 points. The RR intervals that are more than 20\% different than the previous point are considered as artifacts \cite{Cinaz:2013:MMW:2434601.2434680}. RR interval artifact data points are removed. The remaining RR intervals are sampled with cubic spline interpolation for further analysis. After the interpolation, the windows containing more than 10\% interpolated RR interval artifacts are excluded from the model. 20 windows are extracted for each session.
\subsection{Feature Extraction}
For each window, 11 HRV \cite{Alberdi201649} (please see HRV guidelines for further references \cite{hrvguideline}) and 8 acceleration features \cite{TangGSRActivity2014} shown in Table \ref{table:hrv_fatures} are calculated. We extracted features from both time and frequency domains which are determined as the most discriminative ones \cite{survey} by using our MATLAB functions along with Vollmer toolbox \cite{MarcusVollmer}.

We calculated the energy of the acceleration by applying FFT. The mean acceleration over each of the axes is calculated as a feature. A total of 19 features are extracted from the RR intervals coming from the heart rate monitoring unit and acceleration signals. The features and their descriptions are shown in Table \ref{table:hrv_fatures}.

\subsection{Feature Selection and Dimensionality Reduction}

Feature selection is very important for the sake of the machine learning classifiers' performance. In this section, we present the feature selection and dimensionality reduction techniques that we used in our study.

\subsubsection{Correlation Based Feature selection}
We use Correlation Based Feature (CFB) selection. We eliminate highly correlated features. We applied the CFB implementation of the Weka toolkit \cite{weka}. CFB selected 5 features from 16 features. The selected features are RMSSD, SD/SD, mean acceleration over \textit{x} axis, mean acceleration over \textit{z} axis and FFT energy over the acceleration magnitude. 

\subsubsection{Principal Component Analysis}
Principal Component Analysis (PCA) is a dimensionality reduction technique that transforms variables into uncorrelated principal components. It is widely used in data analysis, previous studies reported that when it is used with machine learning algorithms, their classification performance is enhanced. We selected the covered variance as 80\% in order to avoid over-fitting. PCA is used along with the Support Vector Machine.
\subsection{Classification}

In order to discriminate high and low perceived workload level sessions, we use traditional machine learning classifiers in the Weka Toolkit \cite{weka} and LSTM in Keras \cite{chollet2015keras} which is a high level machine learning library which is written in Python and runs on Tensorflow. The model is trained and evaluated on an NVIDIA DGX-1 cluster \cite{dgx}. We created a general model which is trained with all of the participants by not feeding their personal information, i.e., model do not know who is who. Some works created a personal model \cite{Schaule2018} for each participant. Since deep learning requires more data than traditional classifiers, we feed the data coming from participants into a classifier, this approach is known as a general model in the literature. We applied 10-fold stratified cross-validation where all folds class distribution is equal. The fined tuned parameters for the classifiers are as follows:
\begin{itemize}
    \item Random Forest  with 100 Tress
    \item SVM with Radial Basis Function (RBF) kernel (SVM-RBF)
    \item kNN where N=3.
    \item Naive Bayes
    \item Multilayer Perceptron (MLP) with 1 hidden layer and 5 hidden units.
    \item LSTM 200 neurons and 0.2 dropout
\end{itemize}
\begin{figure*}
    \centering
    \includegraphics[width=0.6\columnwidth]{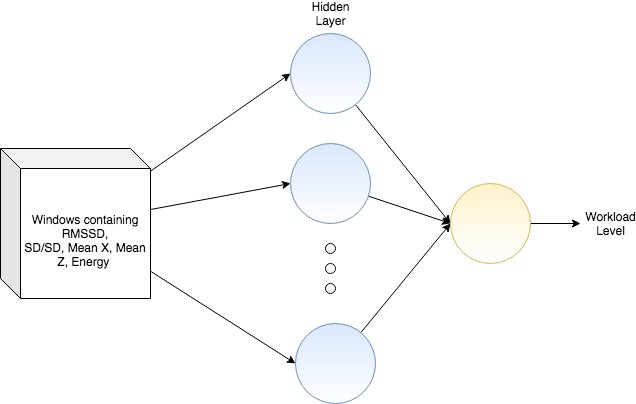}
    \caption{Multilayer Perceptron with one hidden layer and four hidden units. }
    \label{fig:mlp}
\end{figure*}

\begin{figure}
\centering

\includegraphics[ height=2.0in]{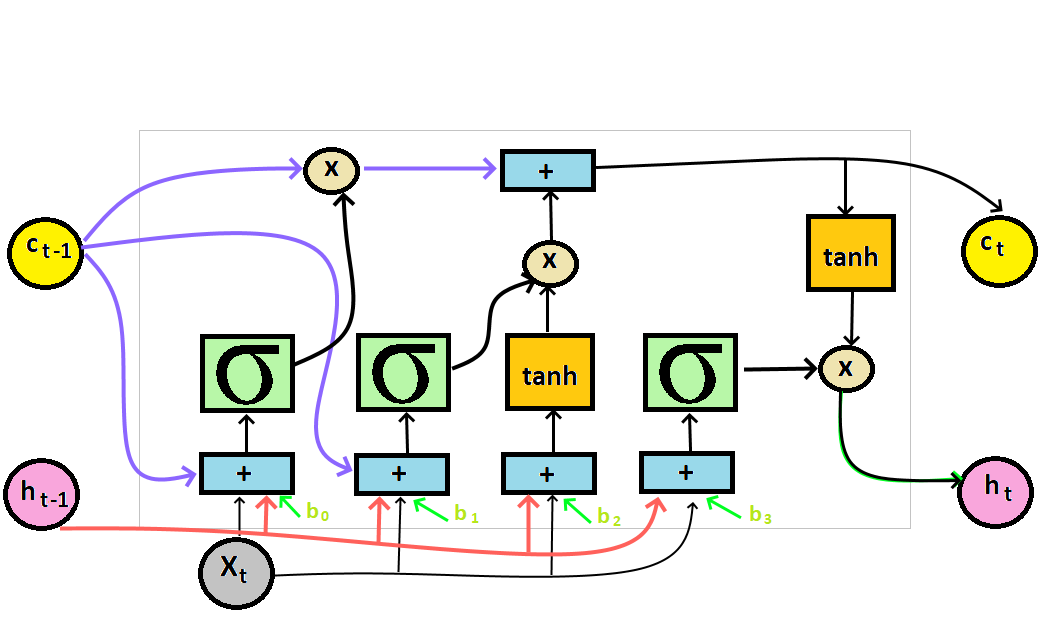}
\caption{The block diagram of an LSTM cell. $\sigma$ is a sigmoid activation function which maps numbers to a range between 0 and 1. Tanh is a hyperbolic tangent activation
function which maps numbers to a range between -1 and 1. $C_t$ is the memory of block t, $H_t$ is the output of block t, input data is the x, past output (block t-1) is $H_{t−1}$, past memory is $C_{t-1}$ and bias vectors are represented as $b_0, b_1, b_2, b_3$ symbols.}
\label{fig:lstm}
\end{figure}

\begin{figure}
\centering

\includegraphics[ width = 0.8\columnwidth]{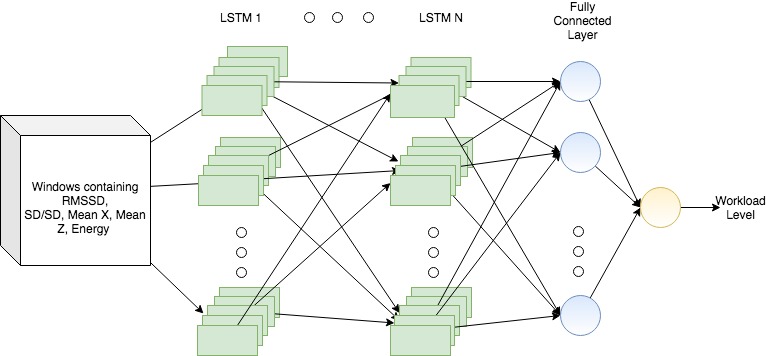}
\caption{The block diagram of the LSTM network that we used in our proposed system. Input is the feature vectors computed from sliding windows.}
\label{fig:lstmour}
\end{figure}

MLP is a feedforward artificial neural network \cite{Alpaydin:2010:IML:1734076}. It has a minimum of three layers which  are the input layer, hidden layer(s) and the output layer. The hidden layer (s) uses the activation functions to capture nonlinear  data relations. Therefore, MLPs can discriminate between classes that are nonlinearly separated \cite{Alpaydin:2010:IML:1734076}. We selected them as a representative of a shallow neural network to compare with the used sequential deep learning method. Unipolar Sigmoid Function was used as an activation function in hidden volumes of MLP. We used a shallow MLP with an only one hidden layer to compare a shallow neural network with the deep sequential one.  The MLP used in our system is shown in Figure  \ref{fig:mlp}.
\\
\indent A basic structure of an LSTM cell is demonstrated in Figure \ref{fig:lstm}. The decision of whether the information should be remembered or not is controlled by gates. The previous 
data is saved via LSTM cells.  There are three types of gates LSTM uses to decide whether to add or remove the information 
in the cell state($C_t$). Forget gate decides to throw away information by using a sigmoid layer. The input gate is the second gate which selects the values to be  updated. It uses a sigmoid layer when making a decision and tanh layer for creating an updated value vector (see Figure \ref{fig:lstm}). The cell state is updated with the reevaluated output of the input gate. The decision of which parts of the cell state is selected as the final output is calculated on the updated cell state and by using a sigmoid layer as seen in Figure \ref{fig:lstm}. The LSTM network that we used in our proposed system is shown in Figure \ref{fig:lstmour}.

For the LSTM networks, we evaluated different number of neurons such as 50, 100, 150, 200 and 250. We also evaluated the effect of the amount of recurrent dropout by applying 0.0, 0.2, 0.4 and 0.6. The fully  connected layer is selected as 25. We also evaluated the LSTM and MLP with different numbers of epochs and we selected 500 as the maximum number of epochs. We will present the best results in terms of the number of epochs.
\section{Experiment Design}
\label{sec:results}
This section evaluates the performance of the proposed daily life perceived workload detection method with traditional classifiers and the LSTM on daily life data obtained from participants.
\begin{figure}
    \centering
    \includegraphics[width=0.2\columnwidth]{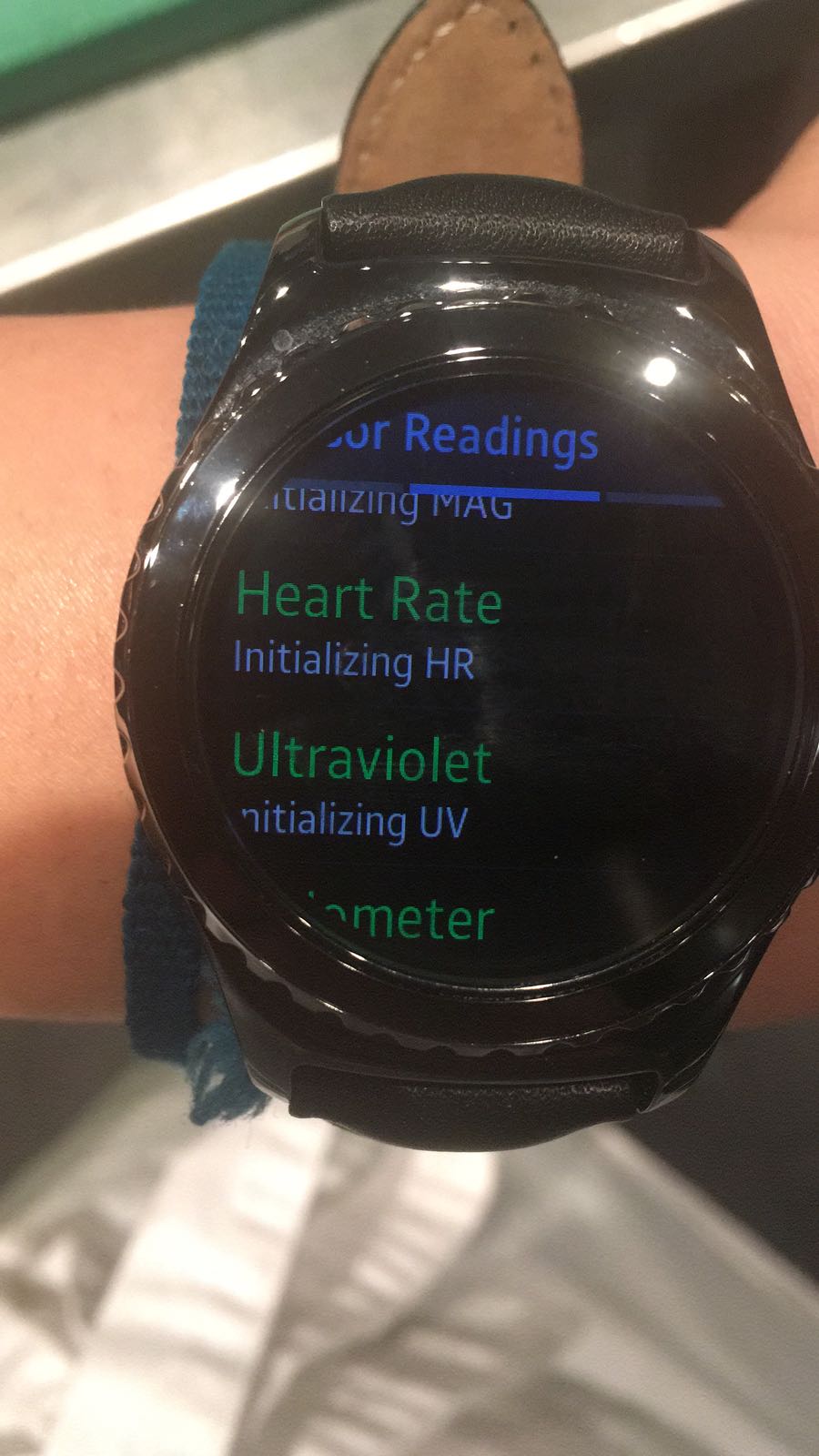}
    \caption{The data collection application developed for the Samsung Gear S2 Smartwatch in Tizen Web Framework.}
    \label{fig:datacollection}
\end{figure}
\subsection{Experiment Design}
\begin{figure*}
\centering
\includegraphics[width=\columnwidth]{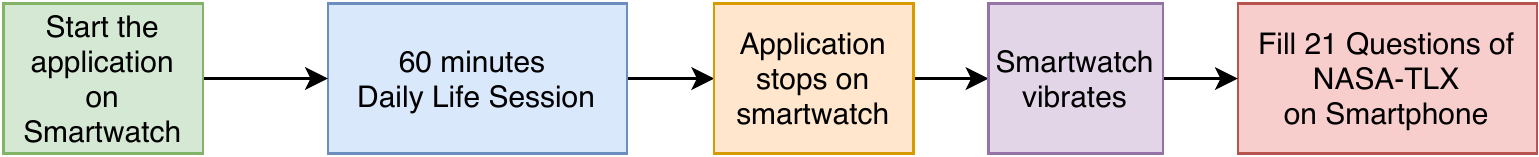}
\caption{The experiment procedure in non-restricted everyday life environment.}
\label{datamethod}
\end{figure*}

\begin{figure*}

    \centering

    \includegraphics[width=0.35\textwidth]{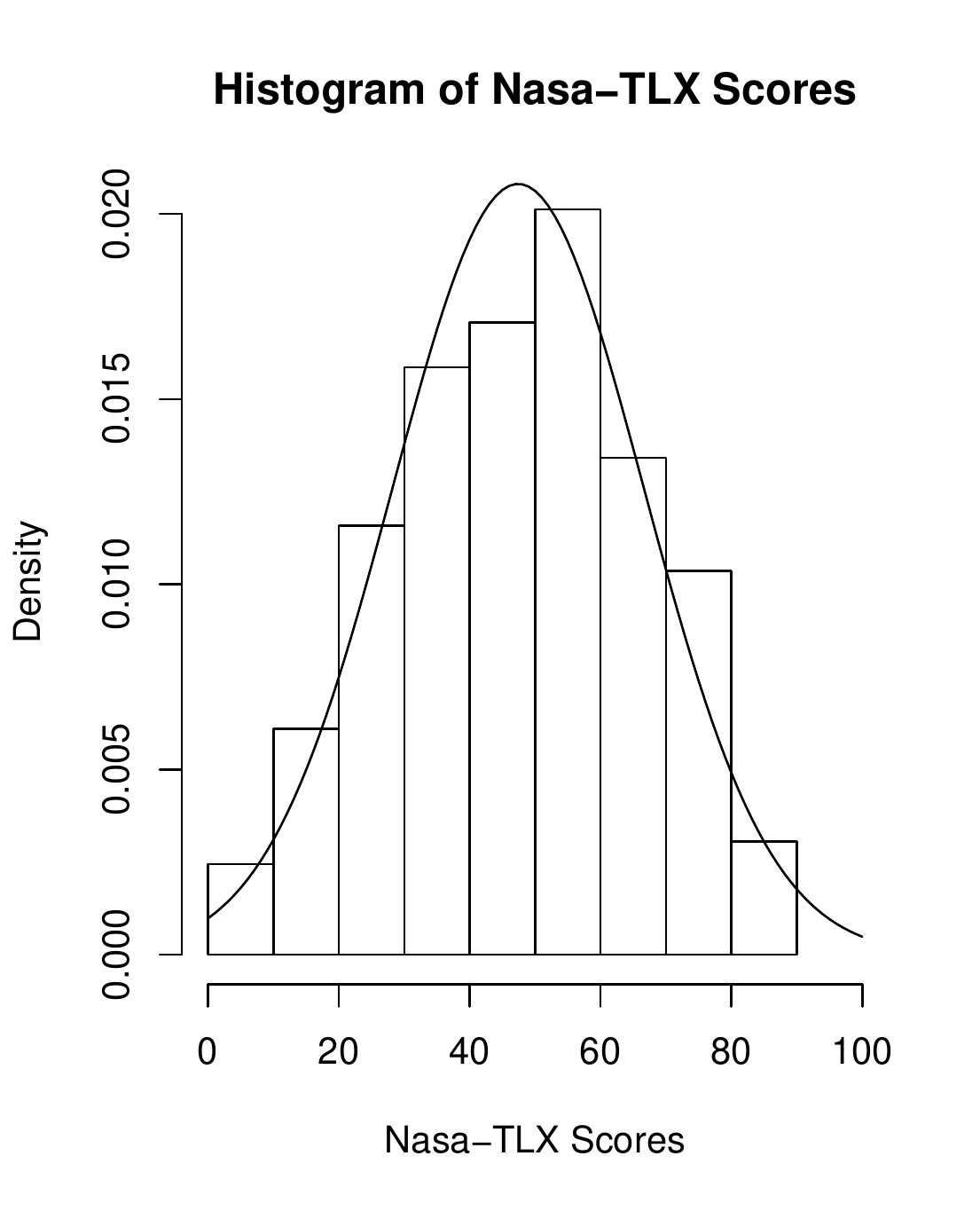}
            \caption{Histogram of NASA-TLX scores.}

    \label{fig:disttlx}
\end{figure*}

Prior to the data acquisition, the user receives an informed consent form. Once the user accepts the terms in the informed consent form, he/she can use the application. The data collected using accelerometer and PPG sensors and stored in the memory of the smartwatch.  Answers of 7854 questions from the NASA Task Load Index (NASA-TLX) questionnaire and 374 hours of physiological signals are collected from 17 participants.  The ages were between 23 and 32. The gender distribution is 13 men and 4 women. Since 5 of the participants did not fill the questionnaires regularly, we could not use their data for supervised learning algorithms. Our early users gave us complaints that they were forgetting to fill the questionnaire, to solve this issue, we added the haptic feedback feature in the next versions of our application and the collaboration of the users for filling the questionnaire after the physiological data collection increased. In Figure \ref{fig:datacollection}, a Samsung Gear S2 smartwatch running our application is shown.  The data is collected over a month from each participant. The resulting amount of questions became 5544 and the duration  of the total recording time became 264 hours. Participants were asked to wear a Samsung Gear S2 smartwatch during their everyday life without any restriction. The procedure of the data collection session in non-restricted everyday life is described in Figure \ref{datamethod}. Participants recorded their physiological signals in their daily life with the application that we developed for Samsung Gear S2. We wanted to make the dataset as rich as possible. Evaluating the physiological data of a participant only on the same day may create a bias. In order to discard this bias, participants are allowed to start our application once a day for an hour, i.e., one trial per day per participant is collected. The length of a trial is 60 minutes. At the end of the session, the participant has received a strong vibration signal from the smartwatch and received a NASA-TLX questionnaire containing 21 questions with 6 scales. The perceived workload score is determined with the NASA-TLX questionnaire (see Table \ref{figure:tlx}). Each participant completed 25 sessions. Generally, participants completed the experiment in a month. The trials with the answers of questionnaires where the time-span between the daily life session and the completion of NASA-TLX are more than 30 minutes and sessions where the data quality is less than 50\% quality are not used for supervised learning.   Hence, we could use 3444 questions and 164 hours of sensory data for the rest of the system. The histogram of the remaining NASA-TLX scores (N=164) is shown in Figure \ref{fig:disttlx}.

\begin{table}[]
\centering
\caption{NASA-TLX factors, rating scales and questions.}
\begin{tabular}{@{}p{4cm}cp{6cm}@{}}
\toprule
\textbf{Factors} & \textbf{ Rating Scale} & \textbf{Questions}   \\ \midrule
Mental Demand               & Low - High   & How much mental and perceptual activity did you spend for this task?       \\
Physical Demand             & Low - High   & How much physical activity did you spend for this task?                         \\
Temporal Demand             & Low - High   & How much time pressure did you feel in order to complete this task?                    \\
Performance                 & Good - Poor  & How successful do you think you were in accomplishing the goals of the task          \\
Effort                      & Low - High   & How hard did you have to work to accomplish your level of performance?                 \\
Frustration                 & Low - High   & How insecure, discouraged, irritated, stressed, and annoyed were you during this task? \\ \bottomrule
\end{tabular}
\label{figure:tlx}
\end{table}
\subsubsection{Collection of Self-Reports}

NASA-TLX \cite{nasatlx} is used to measure the perceived workload of individuals, developed by Hart and Staveland. First, the subject has to rate 6 items on a scale from 0 to 100 that best indicate his experience in the task. The rating consists of the following items: mental demand, physical demand, temporal demand, own performance, effort, and frustration. Next, pairwise comparison of each scale is demonstrated to the subject who is asked to indicate which of the items represents the most important contributor to the stress. Based on these ratings, the total perceived workload was computed as a weighted average. This makes NASA-TLX suitable for measuring the perceived workload in different types of tasks. NASA-TLX can be implemented on a mobile phone \cite{tlxiphone}, paper \cite{nasatlx} or a computer \cite{tlxcomputer1},  \cite{tlxcomputer2}. The six rating scales, questions and endpoints of the mobile implementation of the NASA-TLX are shown in Table \ref{figure:tlx}. In this study, we used the official application of the NASA-TLX \cite{tlxiphone} for Smartphones. NASA-TLX can also be implemented on the smartwatch framework, however the screen sizes of the current smartwatches make it difficult to fill the questionnaire for participants. We recorded self-report questionnaires at the end of each hour and label each session according to their score. The evaluation of self-reports are still an open question.  Some of the models require laboratory calibration session for each individual \cite{cinaz}, \cite{Gjoreski:2016:CSD:2968219.2968306}. The calibration approach makes the enrollment procedure harder and badly affects the scalability of the solution. We used the approach proposed by Sano et al. \cite{sano}. This approach does not require such an enrollment procedure. The sessions are sorted according to NASA-TLX scores.   We divided the dataset into two parts, first $50\%$ as high workload (N=82) and last $50\%$ as low workload (N=82) as shown in Figure \ref{fig:disttlx}. Thanks to this approach, we were able to analyze a balanced dataset in terms of class distribution.
\subsubsection{Ethics}

The procedure of the methodology used in this study is approved by the Institutional Review Board for Research with Human Subjects of Bogazici University with the approval number 2018/16. Prior to the data acquisition, each participant received a consent form which explains the experimental procedure and its benefits and implications to both the society and the subject. The procedure was also explained vocally to the subject. All of the data are stored anonymously.

\section{Results}

The effect of the selection of the LSTM parameters is shown in Table \ref{lstm:tune}. We applied the recurrent dropout, which is available in the Keras implementation, to our LSTM network. The recurrent dropout is a special type of a dropout developed for LSTMs \cite{Semeniuta}, to prevent overfitting for long-term connections.  We evaluated the number of neurons from 50 to 250  and the amount of recurrent dropout from 0.0 to 0.6. The results show that the selection of parameters has a significant impact on the classification accuracy. We compared the results of the LSTM with the MLP and traditional machine learning algorithms in Table \ref{accuracy}. The LSTM network outperformed the traditional machine learning classifiers.

\begin{table}[!htb]
\caption{The effect of different number of neurons and dropout parameters of LSTM networks on classification performance.}
\label{lstm:tune}
\centering
\begin{tabular}{@{}lllll@{}}
\toprule
\textbf{LSTM Blocks/Dropout} & 0.00 & 0.20     & 0.40     & 0.60     \\ \midrule
50 LSTM Blocks     & 67.50\%         & 67.50\% & 65.62\% & 64.37\% \\
100 LSTM Blocks     & 66.87\%         & 66.87\% & 65.62\% & 65.62\% \\
150 LSTM Blocks     & 65.62\%         & 67.50\% & 65.62\% & 63.12\% \\
200 LSTM Blocks     & 65.62\%         & \textbf{70.00\%} & 65.00\% & 64.37\% \\
250 LSTM Blocks     & 66.25\%         & 68.12\% & 66.87\% & 63.75\% \\ \bottomrule
\end{tabular}
\end{table}

\begin{table}[!htb]
\caption{Performance of different classification algorithms. }
\centering
\begin{tabular}{@{}ll@{}}
\toprule
\textbf{Algorithm} & \textbf{Accuracy} \\ \midrule
Random Forest      & 57.92\%           \\
PCA + Support Vector Machine (RBF)     & 65.24\%          \\
K-Nearest neighbor (N=3)         & 57.92\%        \\
Naive Bayes        & 62.80\%         \\
Multi Layer Perceptron       & 60.98\%          \\
Long Short-Term Memory Network               & \textbf{70.00\%}         \\ \bottomrule
\end{tabular}
\label{accuracy}
\end{table}

Monitoring the perceived workload in the wild is not easy, the best performing model is the PCA+SVM (RBF) with 65.24\% accuracy.   The LSTM improved the performance significantly by achieving 70.00 \% accuracy, therefore the application of the RNN is very promising for the unobtrusive and seamless monitoring of the workload in the wild. 
\section{Conclusions}
\label{discussion}
We proposed a perceived workload level monitoring system to be used in the wild settings.  Workload is one of the sources that might lead to negative impact on an individual’s well-being and psychology and that is why it should be monitored closely.  The system trained with self-reports which makes it easily applicable as a smartwatch and smartphone application for the users who want to feed the dataset. We applied the state-of-the-art preprocessing, feature extraction, feature selection, machine learning, and deep learning techniques.  Our system runs on a commercial smartwatch that has rich functionalities. The choice of the device is important because most of the studies use wearables like full-lead ECG or EEG, which are not suitable for the daily life usage due to low comfort, obtrusiveness, and high price. Smartwatches are already part of our daily lives and users do not see them as a burden or obtrusive. This eases the way to adapt our system to daily lives.  However, with these devices, preprocessing and classification techniques of the proposed system gain importance due to the artifacts occur due to motion, a system lacking proper building blocks can easily fail in daily life settings.

\indent Systems that work on unobtrusive consumer-grade smartwatches in the wild would face several additional issues. We tried to find solutions to these problems. The devices might not be properly placed on the subjects, we used the quality measurement of RR intervals and gave visual feedback to the users. Collecting questionnaire data can be hard in daily life settings, a user may forget to fill it due to their busy schedule. We applied haptic feedback thanks to the vibration feature of the smartwatches. This property helped our participants to not forget the filling of the questionnaires. Due to the subjectivity of the questionnaire a model might not fit easily like in the laboratory settings. We tried to solve this issue by conducting a longer period of data collection like a month.

\indent  The supervised methods are evaluated on 164 hours of daily life data and 3444 answers of self-report questions. As seen from the results, LSTM increased the performance of our system approximately 5\% when compared to the best performing classifier.  We demonstrated the importance of the parameter tuning and the application of recurrent dropout for LSTM. Choosing the right parameters in LSTM improves the accuracy of our daily perceived workload detection system.  The results showed that LSTM by its nature fits better with the workload detection problem by using sequential physiological data.  We achieved lower accuracy (70\% vs 90\%) compared to the works carried out in the laboratory. Hence, there is still significant room for improvement for the systems working in the wild settings.  The effect of the smartwatch placement, and tightness of the watch strap should also be evaluated to inform the users for the best usage settings. Even with the haptic feedback of the smartwatch, missing entries can be present. For example, a subject can be in a important meeting where he/she can not reply the questions of the questionnaire. Missing entries of questionnaires with the physiological data can be corrected and used with unsupervised approaches. We believe that the contribution of this study would be useful
especially for students and workers. In particular, it can be potentially beneficial for people who usually experience high workload for a long period of time. By monitoring the workload levels with our unobtrusive daily life applicable system, the productivity of employees could be improved. For instance, if a company detects that a factory worker, pilot, truck driver or soldier experiences enduring high workload levels, they could be given a rest and come back to work as fresh. In this way, possible accidents, negligence could be avoided and the productivity of workers could be improved.





\bibliographystyle{unsrtnat}

\bibliography{cas-sc-template}

\begin{thebibliography}{38}
\providecommand{\natexlab}[1]{#1}
\providecommand{\url}[1]{\texttt{#1}}
\expandafter\ifx\csname urlstyle\endcsname\relax
  \providecommand{\doi}[1]{doi: #1}\else
  \providecommand{\doi}{doi: \begingroup \urlstyle{rm}\Url}\fi

\bibitem[Poungponsri and Yu(2013)]{Poungponsri2013AnAF}
Suranai Poungponsri and Xiao-Hua Yu.
\newblock An adaptive filtering approach for electrocardiogram (ecg) signal
  noise reduction using neural networks.
\newblock \emph{Neurocomputing}, 117:\penalty0 206--213, 2013.

\bibitem[Glaser et~al.(1999)Glaser, Tatum, Nebeker, Sorenson, and
  Aiello]{stressworkload}
Dale~N. Glaser, B.~Charles Tatum, Delbert~M. Nebeker, Richard~C. Sorenson, and
  John~R. Aiello.
\newblock Workload and social support: Effects on performance and stress.
\newblock \emph{Human Performance}, 12\penalty0 (2):\penalty0 155--176, 1999.

\bibitem[Bayram and Bilgel(2008)]{Bayram2008}
Nuran Bayram and Nazan Bilgel.
\newblock The prevalence and socio-demographic correlations of depression,
  anxiety and stress among a group of university students.
\newblock \emph{Social Psychiatry and Psychiatric Epidemiology}, 43\penalty0
  (8):\penalty0 667--672, Aug 2008.
\newblock ISSN 1433-9285.

\bibitem[Rahhal et~al.(2016)Rahhal, Bazi, AlHichri, Alajlan, Melgani, and
  Yager]{RAHHAL2016340}
M.M.~Al Rahhal, Yakoub Bazi, Haikel AlHichri, Naif Alajlan, Farid Melgani, and
  R.R. Yager.
\newblock Deep learning approach for active classification of electrocardiogram
  signals.
\newblock \emph{Information Sciences}, 345:\penalty0 340 -- 354, 2016.

\bibitem[Fan et~al.(2018)Fan, Wade, Key, Warren, and Sarkar]{Fan2018}
Jing Fan, Joshua~W. Wade, Alexandra~P. Key, Zachary~E. Warren, and Nilanjan
  Sarkar.
\newblock {EEG}-based affect and workload recognition in a virtual driving
  environment for {ASD} intervention.
\newblock \emph{{IEEE} Transactions on Biomedical Engineering}, 65\penalty0
  (1):\penalty0 43--51, January 2018.

\bibitem[Can et~al.(2019{\natexlab{a}})Can, Chalabianloo, Ekiz, and
  Ersoy]{Can2019}
Yekta~Said Can, Niaz Chalabianloo, Deniz Ekiz, and Cem Ersoy.
\newblock Continuous stress detection using wearable sensors in real life:
  Algorithmic programming contest case study.
\newblock \emph{Sensors}, 19\penalty0 (8):\penalty0 1849, April
  2019{\natexlab{a}}.

\bibitem[Hochreiter and Schmidhuber(1997)]{hochreiter1997long}
Sepp Hochreiter and J{\"u}rgen Schmidhuber.
\newblock Long short-term memory.
\newblock \emph{Neural computation}, 9\penalty0 (8):\penalty0 1735--1780, 1997.

\bibitem[Alhagry et~al.(2017)Alhagry, Fahmy, and El-Khoribi]{Alhagry2017}
Salma Alhagry, Aly~Aly Fahmy, and Reda~A. El-Khoribi.
\newblock Emotion recognition based on eeg using lstm recurrent neural network.
\newblock \emph{International Journal of Advanced Computer Science and
  Applications}, 8\penalty0 (10), 2017.

\bibitem[Chauhan et~al.(2018)Chauhan, Seneviratne, Hu, Misra, Seneviratne, and
  Lee]{Chauhan2018}
Jagmohan Chauhan, Suranga Seneviratne, Yining Hu, Archan Misra, Aruna
  Seneviratne, and Youngki Lee.
\newblock Breathing-based authentication on resource-constrained {IoT} devices
  using recurrent neural networks.
\newblock \emph{Computer}, 51\penalty0 (5):\penalty0 60--67, May 2018.

\bibitem[Picard(2016)]{PicardLab2016}
R.~W. Picard.
\newblock Automating the recognition of stress and emotion: From lab to
  real-world impact.
\newblock \emph{IEEE MultiMedia}, 23\penalty0 (3):\penalty0 3--7, July 2016.

\bibitem[Cinaz et~al.(2013{\natexlab{a}})Cinaz, Arnrich, Marca, and
  Tr\"{o}ster]{Cinaz:2013:MMW:2434601.2434680}
Burcu Cinaz, Bert Arnrich, Roberto Marca, and Gerhard Tr\"{o}ster.
\newblock Monitoring of mental workload levels during an everyday life
  office-work scenario.
\newblock \emph{Personal Ubiquitous Comput.}, 17\penalty0 (2):\penalty0
  229--239, February 2013{\natexlab{a}}.

\bibitem[Schaule et~al.(2018)Schaule, Johanssen, Bruegge, and
  Loftness]{Schaule2018}
Florian Schaule, Jan~Ole Johanssen, Bernd Bruegge, and Vivian Loftness.
\newblock Employing consumer wearables to detect office
  workers{\textquotesingle} cognitive load for interruption management.
\newblock \emph{Proceedings of the {ACM} on Interactive, Mobile, Wearable and
  Ubiquitous Technologies}, 2\penalty0 (1):\penalty0 1--20, March 2018.

\bibitem[Kane and Conway(2016)]{nback}
Michael Kane and Andrew Conway.
\newblock The invention of n-back: An extremely brief history.
\newblock \emph{The Winnower}, 06 2016.

\bibitem[Munoz et~al.(2016)Munoz, Pereira, and Karapanos]{Munoz2016}
John~Edison Munoz, Fabio Pereira, and Evangelos Karapanos.
\newblock Workload management through glanceable feedback: The role of heart
  rate variability.
\newblock In \emph{2016 {IEEE} 18th International Conference on e-Health
  Networking, Applications and Services (Healthcom)}. {IEEE}, September 2016.

\bibitem[Can et~al.(2019{\natexlab{b}})Can, Arnrich, and Ersoy]{survey}
Yekta~Said Can, Bert Arnrich, and Cem Ersoy.
\newblock Stress detection in daily life scenarios using smart phones and
  wearable sensors: A survey.
\newblock \emph{Journal of Biomedical Informatics}, 92:\penalty0 103139,
  2019{\natexlab{b}}.

\bibitem[{Ram} et~al.(2012){Ram}, {Madhav}, {Krishna}, {Komalla}, and
  {Reddy}]{motionppg}
M.~R. {Ram}, K.~V. {Madhav}, E.~H. {Krishna}, N.~R. {Komalla}, and K.~A.
  {Reddy}.
\newblock A novel approach for motion artifact reduction in ppg signals based
  on as-lms adaptive filter.
\newblock \emph{IEEE Transactions on Instrumentation and Measurement},
  61\penalty0 (5):\penalty0 1445--1457, May 2012.

\bibitem[Ekiz et~al.(2017)Ekiz, Kaya, Bugur, Guler, Buz, Kosucu, and
  Arnrich]{Ekiz2017}
Deniz Ekiz, Gamze~Ege Kaya, Serkan Bugur, Sila Guler, Buse Buz, Bilgin Kosucu,
  and Bert Arnrich.
\newblock Sign sentence recognition with smart watches.
\newblock In \emph{2017 25th Signal Processing and Communications Applications
  Conference ({SIU})}. {IEEE}, May 2017.

\bibitem[Chalabianloo et~al.(2018)Chalabianloo, Ekiz, Can, and Ersoy]{hibit}
Niaz Chalabianloo, Deniz Ekiz, Yekta~Said Can, and Cem Ersoy.
\newblock Smart watch based stress detection in real life.
\newblock In \emph{11th International Symposium on Health Informatics and
  Bioinformatics}, page~39, Antalya, 2018.

\bibitem[Ekman(1965)]{Ekman1965}
Paul Ekman.
\newblock Differential communication of affect by head and body cues.
\newblock \emph{Journal of Personality and Social Psychology}, 2\penalty0
  (5):\penalty0 726--735, 1965.

\bibitem[Montepare et~al.(1999)Montepare, Koff, Zaitchik, and
  Albert]{Montepare1999}
Joann Montepare, Elissa Koff, Deborah Zaitchik, and Marilyn Albert.
\newblock \emph{Journal of Nonverbal Behavior}, 23\penalty0 (2):\penalty0
  133--152, 1999.

\bibitem[Mat(2015)]{matlab}
\emph{MATLAB version 8.5.0.197613 (R2015a)}.
\newblock The Mathworks, Inc., Natick, Massachusetts, 2015.

\bibitem[Gjoreski et~al.(2017)Gjoreski, Lu{\v{s}}trek, Gams, and
  Gjoreski]{gjoreski2017monitoring}
Martin Gjoreski, Mitja Lu{\v{s}}trek, Matja{\v{z}} Gams, and Hristijan
  Gjoreski.
\newblock Monitoring stress with a wrist device using context.
\newblock \emph{Journal of biomedical informatics}, 73:\penalty0 159--170,
  2017.

\bibitem[Alberdi et~al.(2016)Alberdi, Aztiria, and Basarab]{Alberdi201649}
Ane Alberdi, Asier Aztiria, and Adrian Basarab.
\newblock Towards an automatic early stress recognition system for office
  environments based on multimodal measurements: A review.
\newblock \emph{Journal of Biomedical Informatics}, 59:\penalty0 49 -- 75,
  2016.

\bibitem[of~The European Society~of Cardiology et~al.(1996)of~The European
  Society~of Cardiology, of~Pacing, and Electrophysiology]{hrvguideline}
Task~Force of~The European Society~of Cardiology, The North American~Society
  of~Pacing, and Electrophysiology.
\newblock Heart rate variability, standards of measurement, physiological
  interpretation, and clinical use.
\newblock \emph{European Heart Journal}, 17:\penalty0 354--381, 1996.

\bibitem[Tang et~al.(2014)Tang, Yeo, and Lau]{TangGSRActivity2014}
T.~B. Tang, L.~W. Yeo, and D.~J.~H. Lau.
\newblock Activity awareness can improve continuous stress detection in
  galvanic skin response.
\newblock In \emph{IEEE SENSORS 2014 Proceedings}, pages 1980--1983, Nov 2014.

\bibitem[Vollmer()]{MarcusVollmer}
Marcus Vollmer.
\newblock Marcusvollmer/hrv toolbox.
\newblock URL \url{https://www.github.com/MarcusVollmer/HRV Retrieved October
  24, 2018}.

\bibitem[Holmes et~al.(1994)Holmes, Donkin, and Witten]{weka}
G.~Holmes, A.~Donkin, and I.~H. Witten.
\newblock Weka: a machine learning workbench.
\newblock In \emph{Proceedings of ANZIIS '94 - Australian New Zealnd
  Intelligent Information Systems Conference}, pages 357--361, Nov 1994.

\bibitem[Chollet et~al.(2015)]{chollet2015keras}
Fran\c{c}ois Chollet et~al.
\newblock Keras.
\newblock \url{https://keras.io}, 2015.
\newblock {accessed at November 2019}.

\bibitem[dgx(2019)]{dgx}
\emph{AI Infrastructure Deployment with DGX-1}, 2019.
\newblock URL
  \url{https://www.nvidia.com/en-us/data-center/dgx-pod-reference-architecture/}.
\newblock {accessed at October 2019}.

\bibitem[Alpaydin(2010)]{Alpaydin:2010:IML:1734076}
Ethem Alpaydin.
\newblock \emph{Introduction to Machine Learning}.
\newblock The MIT Press, 2nd edition, 2010.

\bibitem[Hard and Stavenland(1988)]{nasatlx}
SG~Hard and Stavenland.
\newblock Development of nasa-tlx (task load index): results of empirical and
  theoretical research.
\newblock \emph{Advances in Psychology}, 52:\penalty0 139 -- 183, 1988.

\bibitem[tlx(2018)]{tlxiphone}
Nasa tlx for ios user guide v1.0, 2018.
\newblock URL
  \url{https://humansystems.arc.nasa.gov/groups/TLX/downloads/NASA_TLX_for_iOS_User_Guide_Final.pdf}.
\newblock accessed at November 2019.

\bibitem[Cao et~al.(2009)Cao, Chintamani, Pandya, and Ellis]{tlxcomputer1}
Alex Cao, Keshav~K. Chintamani, Abhilash~K. Pandya, and R.~Darin Ellis.
\newblock Nasa tlx: Software for assessing subjective mental workload.
\newblock \emph{Behavior Research Methods}, 41\penalty0 (1):\penalty0 113--117,
  Feb 2009.
\newblock ISSN 1554-3528.

\bibitem[Sharek(2011)]{tlxcomputer2}
David Sharek.
\newblock A useable, online nasa-tlx tool.
\newblock \emph{Proceedings of the Human Factors and Ergonomics Society Annual
  Meeting}, 55\penalty0 (1):\penalty0 1375--1379, 2011.

\bibitem[Cinaz et~al.(2013{\natexlab{b}})Cinaz, Arnrich, Marca, and
  G.Tröster]{cinaz}
B.~Cinaz, B.~Arnrich, R.~Marca, and G.Tröster.
\newblock Monitoring of mental workload levels during an everyday life
  office-work scenario.
\newblock \emph{Personal Ubiquitous Comput.}, 2013{\natexlab{b}}.

\bibitem[Gjoreski et~al.(2016)Gjoreski, Gjoreski, Lu\v{s}trek, and
  Gams]{Gjoreski:2016:CSD:2968219.2968306}
Martin Gjoreski, Hristijan Gjoreski, Mitja Lu\v{s}trek, and Matja\v{z} Gams.
\newblock Continuous stress detection using a wrist device: In laboratory and
  real life.
\newblock In \emph{Proceedings of the 2016 ACM International Joint Conference
  on Pervasive and Ubiquitous Computing: Adjunct}, UbiComp '16, pages
  1185--1193, New York, NY, USA, 2016. ACM.

\bibitem[Sano et~al.(2018)Sano, Taylor, McHill, Phillips, Barger, Klerman, and
  Picard]{sano}
Akane Sano, Sara Taylor, Andrew~W McHill, Andrew~JK Phillips, Laura~K Barger,
  Elizabeth Klerman, and Rosalind Picard.
\newblock Identifying objective physiological markers and modifiable behaviors
  for self-reported stress and mental health status using wearable sensors and
  mobile phones: Observational study.
\newblock \emph{J Med Internet Res}, 20\penalty0 (6):\penalty0 e210, Jun 2018.
\newblock ISSN 1438-8871.

\bibitem[Semeniuta et~al.(2016)Semeniuta, Severyn, and Barth]{Semeniuta}
Stanislau Semeniuta, Aliaksei Severyn, and Erhardt Barth.
\newblock Recurrent dropout without memory loss.
\newblock \emph{CoRR}, abs/1603.05118, 2016.
\newblock URL \url{http://arxiv.org/abs/1603.05118}.

\end{thebibliography}


\end{document}